\documentclass[10pt,twocolumn]{IEEEtran}
\usepackage{amsmath,amssymb,mathrsfs}
\usepackage{color,array,times}
\usepackage{epsfig,epstopdf,textcomp,graphicx}
\begin{document}

\title{Study of Distributed Robust Beamforming with Low-Rank and Cross-Correlation Techniques}

\author{Hang Ruan and Rodrigo C. de Lamare}

\maketitle

\begin{abstract}
In this work, we present a novel robust distributed beamforming (RDB) approach based on low-rank and cross-correlation techniques. The proposed RDB approach mitigates the effects of channel errors in
wireless networks equipped with relays based on the exploitation of
the cross-correlation between the received data from the relays at
the destination and the system output and low-rank techniques. The
relay nodes are equipped with an amplify-and-forward (AF) protocol
and the channel errors are modeled using an additive matrix
perturbation, which results in degradation of the system
performance. The proposed method, denoted low-rank and
cross-correlation RDB (LRCC-RDB), considers a total relay transmit
power constraint in the system and the goal of maximizing the output
signal-to-interference-plus-noise ratio (SINR). We carry out a
performance analysis of the proposed LRCC-RDB technique along with a
computational complexity study. The proposed LRCC-RDB does not
require any costly online optimization procedure and simulations
show an excellent performance as compared to previously reported
algorithms.

\end{abstract}


\section{Introduction}

Distributed beamforming techniques have been widely investigated in
sensor array signal processing in recent years \cite{r1,r2,r3} along
with their applications. Such techniques can be highly useful for
situations in which the channels between emitting sources and
destination devices have poor quality so that the latter cannot
communicate directly and relies on relays that receive, process and
forward the signals. In this context, low-power devices can
substantially enhance the quality of the received signal and
mitigate interference.

\subsection{Prior and Related Work}

Prior work on distributed beamforming
\cite{r1,r2,r3,r4,r7,r29,r30,r5,r6,r23,r13,zhu2016,ma2017,koutrovelis2018}
includes several approaches to enhancing the reception of signals
originating from relays. In \cite{r2}, relay network problems are
described as optimization problems and relevant aspects and
implications are provided and discussed, which gives a general
overview and methodologies that are commonly considered to
analyze relay networks. The work in \cite{r4} focuses on multiple
scenarios with different optimization problem formulations in order
to optimize the beamforming weight vector and increase the
signal-to-noise ratio (SNR) of the system. The work in \cite{r7} considers an optimization
problem that maximizes the output signal-to-interference-plus-noise
ratio (SINR) under total relay transmit power constraints, by
selecting the beamforming weights in such a way that they can
compute the weights using local information, whereas the selection
procedure still depends on global CSI. Similarly, compared to
\cite{r7}, our previous work in \cite{r29} combines relay selection
and a consensus algorithm to enable the local information at all
relays exchangeable without losing much performance. Moreover, SINR
maximization approaches can also be associated with relay selection
to reduce system complexity as in \cite{r30}.

Some other work focuses on power control and allocation strategies at the relay nodes as
well as the overall system power consumption, rather than the system
performance in terms of output SNR or SINR. The studies in
\cite{r5,r6} analyze power control methods based on channel
magnitude, whereas the powers of each relay are adaptively adjusted
according to the qualities of their associated channels. The study
in \cite{r23} uses joint distributed beamforming and power
allocation in underlay cognitive two-way relay links using
second-order channel statistics for SINR balancing and maximization.
However, most of these approaches are derived by assuming that the
global CSI is perfectly known. The work in \cite{r13} only requires
local CSI but employs a different system model, which employs a
reference signal-based scheme.

However, in most scenarios encountered  the channels observed by the
relays may lead to performance degradation because of inevitable
measurements, estimation procedures and quantization errors in CSI
\cite{r12} as well as propagation effects. These impairments result
in imperfect CSI that can affect most distributed beamforming
methods, which either fail to work properly or cannot provide
satisfactory performance. In this context, robust distributed
beamforming (RDB) techniques are hence in demand to mitigate the
channel errors or uncertainties and preserve the relay system
performance. The studies in \cite{r12,r9,r11,r25} minimize the total
relay transmit power under an overall quality of service (QoS)
constraint, using either a convex semi-definite programme (SDP)
relaxation method or a convex second-order cone programme (SOCP).
The works in \cite{r12,r9} consider the channel errors as Gaussian
random vectors with known statistical distributions between the
source to the relay nodes and the relay nodes to the destination,
whereas \cite{r11} models the channel errors on their covariance
matrices as a type of matrix perturbation. The work in
\cite{r11,r14,r21} presents a robust design, which ensures that the
SNR constraint is satisfied in the presence of imperfect CSI by
adopting a worst-case design and formulates the problem as a convex
optimization problem that can be solved efficiently. Similar
approaches that use the worst-case method can be found in
conventional beamforming as in \cite{r26,r28}. The study in
\cite{r15} discusses multicell coordinated beamforming in the
presence of CSI errors, where base stations (BSs) collaboratively
mitigate their intercell interference (ICI). An optimization problem
that minimizes the overall transmission power subject to multiple
QoS constraints is considered and solved using semi-definite
relaxation (SDR) and the S-Lemma. The work in \cite{r24} studies a
systematic analytic framework for the convergence of a general set
of adaptive schemes and their tracking capability with stochastic
stability. The work in \cite{LZhang2013} discusses different design
and optimization criteria for distributed beamforming problems with
perfect instantaneous CSI as well as low-complexity real-valued
implementations, where a generalized eigenvector problem (GEP) is
solved for SNR maximization in the presence of CSI errors. The study
in \cite{RMudumbai2007} proposes a master-slave architecture to show
that most gains of distributed transmit beamforming can be obtained
with imperfect synchronization corresponding to phase errors with
moderate variance. Similarly, the work in \cite{RMudumbai2010}
devises a distributed adaptation method for the transmitters with
minimal feedback from the receiver to ensure phase coherence of the
radio frequency signals from different transmitters in the presence
of unknown phase offsets between the transmitters and unknown CSI
from the transmitters to the receiver.

\subsection{Contributions}

In this work, we propose an RDB technique that achieves very high
estimation accuracy in terms of channel mismatch with reduced
computational complexity in scenarios where the global CSI is
imperfect and local communication is unavailable. Specifically, we show that the proposed technique is versatile enough to tackle scenarios based on different CSI availability assumptions:
\begin{itemize}
    \item when the instantaneous CSI is available while the CSI second-order statistics is not, the channel covariance matrices are iteratively estimated and then channel error spectrum matrices are iteratively constructed;
    \item when the CSI second-order statistics is available while the instantaneous CSI is not, the channel error spectrum matrices can be directly constructed out of any iteration.
    \end{itemize} Meanwhile, unlike most of the existing RDB approaches, we aim to maximize the system
output SINR subject to a total relay transmit power constraint using
an approach that exploits the cross-correlation between the
beamforming weight vector and the system output and then projects
the obtained cross-correlation vector onto subspace computed from
the channel error spectrum matrices to produce more accurate CSI
estimates, namely, the low-rank and cross-correlation robust
distributed beamforming (LRCC-RDB) technique. Unlike our previous
work on centralized beamforming \cite{r17}, the LRCC-RDB technique
is distributed and has marked differences in the way the subspace
processing is carried out. In the LRCC-RDB approach, the covariance
matrices of the channel errors are modeled by additive matrix
perturbation \cite{r8}, which ensures that the covariance matrices
are always positive-definite. We consider multiple signal sources
and assume that there is no direct link between them and the
destination. We consider that the channel errors exist both between
the signal sources and the relays and between the relays and the
destination. The channel error is decomposed and estimated for each
signal originating from a source at each time instant separately.
The proposed LRCC-RDB technique shows outstanding SINR performance
as compared to the existing distributed beamforming techniques,
which focus on transmit power minimization over a wide range of
system input SNR values.

In summary, the main contributions of this work are:
\begin{itemize}
  \item The proposed LRCC-RDB technique, an iterative approach that delivers highly accurate estimates for the relay channel errors as well as the beamforming weights in order to maximize the system output SINR, when the global CSI is imperfect and a total relay transmit power constraint is imposed.
  \item A comprehensive mean squared error (MSE) analysis of the proposed LRCC-RDB technique as well as a general distributed beamforming scenario with channel errors when no specific RDB technique is applied.
  \item An analysis of computational complexity along with comparisons to the relevant existing RDB techniques.
  \item A simulation study of the proposed LRCC-RDB and existing RDB algorithms in several scenarios of interest.
\end{itemize}

This paper is organized as follows. Section II presents the system
model and states the problem. In Section III, the proposed LRCC-RDB
technique is introduced. Section IV present the performance analysis
and a study of the computational complexity of LRCC-RDB and existing
RDB techniques. Section V presents and discusses the simulation
results. Section VI gives the conclusion.

\subsection{Notation}

The notation adopted in this paper includes: lowercase non-bold
letters represent scalar values whereas bold lowercase and upper
case letters represent vectors and matrices, respectively. $(.)^*$,
$(.)^T$, $(.)^{-1}$ and $(.)^H$ denote the complex conjugate
operator, the transpose operator, matrix inversion operator and the
Hermitian transpose operator, respectively. $|.|$, $||.||$, and
${||.||}_F$ denote the absolutely value of a scalar, the Euclidean
norm of a vector or matrix and the Frobenius norm of a vector or matrix, respectively. The symbol $\odot$ represents the Schur-Hadamard product. $E[.]$ denotes expectation. ${\rm tr}(.)$ and ${\rm diag}(.)$ denote the trace and the diagonal of a matrix, respectively. An identity matrix of size $M$ is represented by ${\bf I}_M$.

\section{System Model and Problem Statement}

\begin{figure}[!htb]
\begin{center}
\def\epsfsize#1#2{0.925\columnwidth}
\epsfbox{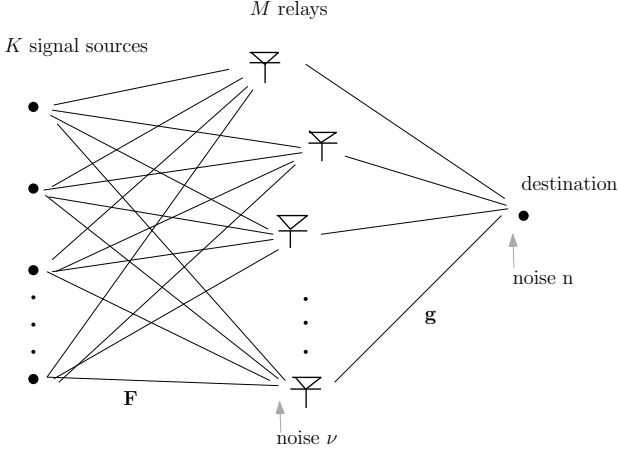} \caption{System model} \label{model}
\end{center}
\end{figure}

We consider a wireless communication network that consists of $K$
signal sources (one desired signal source and $K-1$ interferers),
$M$ distributed single-antenna relays and a destination. It is
assumed that the quality of the channels between the signal sources
and the destination is such that direct communication is not
reliable and their links are negligible. The $M$ relays receive
signals transmitted by the sources and then retransmit them to a
destination by employing a beamforming procedure, in which a
two-step amplify-and-forward (AF) protocol is considered, as shown
in Fig. \ref{model}. We only consider the AF protocol because it
requires lower computing power than other protocols
\cite{jpais,smce,tdscl,tds,armo} which helps to reduce cost and
transmit power as decoding and costly signal processing are not
needed at the relay nodes. Moreover, the AF protocol requires relay
nodes to operate from time-slot to time-slot, which makes it the
most appropriate alternative for the proposed LRCC-RDB algorithm to
work as it relies on iterations over time-slots. Extensions to other
protocols are left for future work.

In the first transmission phase, the $K$ sources transmit the
signals to the $M$ single-antenna relays according to the model
given by
\begin{equation}
{\bf x}={\bf F}{\bf s}+{\boldsymbol \nu}, \label{eq1}
\end{equation}
where the vector ${\bf s}=[s_1, s_2, \dotsb, s_K]^T \in {\mathbb
C}^{K \times 1}$ contains signals with zero mean denoted by
$s_k=\sqrt{P_{s,k}}b_k$ for $k=1,2,\dotsb,K$, where
$P_{s,k}$, $b_k$ and $\sigma^2_{b_k}=E[|b_k|^2]$ are the transmit
power, information symbol and variance of the information symbol of the $k$th signal source, respectively. We assume that $s_1$ is the desired signal while the
remaining signals are treated as interferers. Note that other
configurations with multiple destinations are also possible. The
matrix ${\bf F}=[{\bf f}_1, {\bf f}_2, \dotsb, {\bf f}_K] \in
{\mathbb C}^{M \times K}$ is the channel matrix between the signal
sources and the relays, ${\bf f}_k=[f_{1,k}, f_{2,k}, \dotsb,
f_{M,k}]^T \in {\mathbb C}^{M \times 1}$, $f_{m,k}$ denotes the
channel between the $m$th relay and the $k$th source ($m=1,2,
\dotsb, M$, $k=1,2,\dotsb, K$). ${\boldsymbol \nu}=[\nu_1, \nu_2,
\dotsb, \nu_M]^T \in {\mathbb C}^{M \times 1}$ is the complex
Gaussian noise vector at the relays and $\sigma_{\nu}^2$ is the
noise variance at each relay ($\nu_m$ $\sim$ ${\mathcal
CN}(0,\sigma_{\nu}^2)$ refers to the complex Gaussian distribution
with zero mean and variance $\sigma_{\nu}^2$).

The vector ${\bf x} \in {\mathbb C}^{M \times 1}$ represents the
received signal at the relays. In the second transmission phase, the
relays transmit ${\bf y} \in {\mathbb C}^{M \times 1}$, which is an
amplified and phase-steered version of ${\bf x}$ that can be written
as
\begin{equation}
{\bf y}={\bf W}{\bf x}, \label{eq2}
\end{equation}
where ${\bf W}={\rm diag}([w_1, w_2, \dotsb, w_M]) \in {\mathbb
C}^{M \times M}$ is a diagonal matrix whose entries denote the
beamforming weights. Then the signal received at the destination is
given by
\begin{equation}
z={\bf g}^T{\bf y}+n, \label{eq3}
\end{equation}
where $z$ is a scalar, ${\bf g}=[g_1, g_2, \dotsb, g_M]^T \in
{\mathbb C}^{M \times 1}$ is the complex Gaussian channel vector
between the relays and the destination, $n$ ($n$ $\sim$ ${\mathcal
CN}(0,\sigma_n^2)$) is the noise at the destination and $z$ is the
received signal at the destination. Here we assume for simplicity
that the noise samples at each relay and the destination have the
same power, that is, $P_n=\sigma_n^2=\sigma_{\nu}^2$.

The channel matrices ${\bf F}$ and ${\bf g}$ are modeled as Rayleigh distributed random variables and we also consider distance-based large-scale channel propagation effects such as path loss and shadowing. An exponential based path
loss model is described by \cite{r16}
\begin{equation}
\gamma_m=\frac{\sqrt{L}}{\sqrt{d_m^{\rho}}}, \label{eq4}
\end{equation}
where $\gamma_m$ is the distance-based path loss, $L$ is the known
path loss at the destination, $d_m$ is the distance of interest
relative to the destination and $\rho$ is the path loss exponent,
which can vary due to different environments and is typically set
within $2$ to $5$, with a lower value representing a clear and
uncluttered environment, which has a slow attenuation, and a higher
value describing a cluttered and highly attenuating environment. Shadow fading can be described as a random variable with a
probability distribution described by \cite{r16}
\begin{equation}
\beta_m=10^{(\frac{\sigma_s \eta}{10})}, \label{eq5}
\end{equation}
where $\beta$ is the shadowing parameter, $\eta \sim {\mathcal N}(0,1)$ means $\eta$ is drawn from a Gaussian distribution with zero mean and unit variance and
$\sigma_s$ is the shadowing spread in dB. The shadowing spread
reflects the severity of the long-term attenuation caused by
shadowing, and is typically given between $0$dB to $9$dB \cite{r16}. Without losing generality, for simplicity we assume all relays share the same value of $\rho$ and $\sigma_s$, and that the signal sources are close enough so that they can be treated as a source pool. The distances of the source-to-relay
links $d_{s,r_m}$ ($m=1,\dotsb,M$) are modeled as pseudo-random in
an area defined by a range of relative distances based on the
source-to-destination distance $d_{s,d}$ which is set to $1$, so as
the source-to-relay link distances $d_{s,r_m}$ are decided by a set
of uniform random variables distributed between $0.5$ to $0.9$, with
corresponding relay-source-destination angles $\theta_{r_m,s,d}$
randomly chosen from an angular range of $-\pi/2$ to $\pi/2$.
Therefore, the relay-to-destination distances $d_{r_m,d}$ can be
calculated using the trigonometrical identity given by
$$ d_{r_m,d}=\sqrt{d_{s,r_m}^2+1-2d_{s,r_m}\cos\theta_{r_m,s,d}}.$$ The channels modeled with both path-loss and shadowing can be
represented by
\begin{equation}
{\bf f}_m=\gamma_m\beta_m{\bf f}_{0,m},  \label{eq6}
\end{equation}
\begin{equation}
g_m=\gamma_m\beta_m{g_{0,m}},  \label{eq7}
\end{equation}
where ${\bf f}_{0,m}$ and $g_{0,m}$ denote the Rayleigh distributed channels of the $m$th relay without large-scale propagation effects.

The received signal at the $m$th relay can be expressed as:
\begin{equation}
x_m=\sum_{k=1}^K \underbrace{\sqrt{P_{s,k}}{b_k}}_{s_k}
f_{m,k}+{\nu}_m, \label{eq8}
\end{equation}
then the transmitted signal at the $m$th relay is given by
\begin{equation}
y_m={w_m}x_m. \label{eq9}
\end{equation}
The transmit power at the $m$th relay is equivalent to $E[|y_m|^2]$
so that it can be written as
$\sum_{m=1}^ME[|y_m|^2]=\sum_{m=1}^ME[|{w_m}x_m|^2]$ or in matrix
form as ${\bf w}^H{\bf D}{\bf w}$ where ${\bf D}={\rm
diag}\big(\sum_{k=1}^KP_{s,k}{\sigma^2_{b_k}}\big[E[|f_{1,k}|^2],
E[|f_{2,k}{|}^2], \dotsb, E[|f_{M,k}{|}^2]\big]+P_n\big)$ is a
full-rank matrix. The signal received at the destination can be
expanded by substituting \eqref{eq8} and \eqref{eq9} in \eqref{eq3},
which yields
\begin{multline}
z=\underbrace{\sum_{m=1}^M{w_m}g_m\sqrt{P_{s,1}}f_{m,1}{b_1}}_{\text{desired
signal}}
+\underbrace{\sum_{m=1}^M{w_m}g_m\sum_{k=2}^K\sqrt{P_{s,k}}f_{m,k}{b_k}}_{\text{interferers}}\\
+\underbrace{\sum_{m=1}^M{w_m}g_m\nu_m+n}_{\text{noise}}.
\label{eq10}
\end{multline}
By taking expectation of the components of \eqref{eq10}, we can
compute the desired signal power $P_{z,1}$, the interference power
$P_{z,i}$ and the noise power $P_{z,n}$ at the destination as
follows:
\begin{multline}
P_{z,1}=E\Big[\sum_{m=1}^M({w_m}g_m\sqrt{P_{s,1}}f_{m,1}{b_1})^2\Big] \\
=P_{s,1}{\sigma^2_{b_1}}\underbrace{\sum_{m=1}^ME\Big[w_m^*(f_{m,1}g_m)(f_{m,1}g_m)^*w_m\Big]}_{{{\bf
w}^HE[({\bf f}_1 \odot {\bf g})({\bf f}_1 \odot {\bf g})^H]{\bf
w}}}, \label{eq11}
\end{multline}
\begin{equation}
\begin{split}
P_{z,i}& = E\Big[\sum_{m=1}^M({w_m}g_m\sum_{k=2}^K\sqrt{P_{s,k}}f_{m,k}{b_k})^2\Big] \\
& = \sum_{k=2}^KP_{s,k}{\sigma^2_{b_k}}
\underbrace{\sum_{m=1}^ME\Big[w_m^*(f_{m,k}g_m)(f_{m,k}g_m)^*w_m\Big]}_{{{\bf
w}^HE[({\bf f}_k \odot {\bf g})({\bf f}_k \odot {\bf g})^H]{\bf w}}}
\label{eq12}
\end{split}
\end{equation}
\begin{equation}
\begin{split}
P_{z,n}& =E\Big[\sum_{m=1}^M({w_m}g_m\nu_m+n)^2\Big] \\
& =P_n(1+\underbrace{\sum_{m=1}^ME\Big[w_m^*g_mg_m^*w_m\Big]}_{{{\bf
w}^HE[{\bf g}{\bf g}^H]{\bf w}}}), \label{eq13}
\end{split}
\end{equation}
where $*$ denotes complex conjugate. By defining $${\bf R}_k
\triangleq P_{s,k}{\sigma^2_{b_k}}E[({\bf f}_k \odot {\bf g})({\bf
f}_k \odot {\bf g})^H],$$ for $k=1,2,\dotsb,K$ and
$${\bf Q} \triangleq P_nE[{\bf g}{\bf g}^H],$$ the SINR is computed
as:
\begin{equation}
\begin{split}
{\rm SINR} &  =\frac{P_{z,1}}{P_{z,i}+P_{z,n}} = \frac{{\bf w}^H{\bf
R}_1{\bf w}}{P_n +{\bf w}^H({\bf Q}+\sum_{k=2}^K{\bf R}_k){\bf w}}.
\label{eq14}
\end{split}
\end{equation}
Note that in \eqref{eq14} the quantities ${\bf R}_k$, $k=1,\dotsb,K$
and ${\bf Q}$ only consist of the second-order statistics of the
channels, which means the channels have no mismatches and they
correspond to perfect CSI knowledge. In order to consider errors in
the channels $\hat{\bf F}$ and $\hat{\bf g}$, we introduce the
matrix ${\bf E}=[{\bf e}_1,\dotsb,{\bf e}_K] \in {\mathbb C}^{M
\times K}$ and the vector ${\bf e} \in {\mathbb C}^{M \times 1}$,
which yield
\begin{equation}
\hat{\bf f}_k={\bf f}_k+{\bf e}_k,  k=1,2,\dotsb,K,  \label{eq15}
\end{equation}
\begin{equation}
\hat{\bf g}={\bf g}+{\bf e},  k=1,2,\dotsb,K,  \label{eq15+}
\end{equation}
where $\hat{\bf f}_k$ is the $k$th mismatched channel component of
${\bf F}$. The elements of ${\bf E}$, i.e., ${\bf e}_k$ for any
$k=1,\dotsb,K$ and ${\bf e}$, are assumed to be for simplicity
independent and identically distributed (i.i.d) Gaussian variables
so that the covariance matrices ${\bf R}_{{\bf e}_k}=E[{\bf e}_k{\bf
e}_k^H]$ and ${\bf R}_{\bf e}=E[{\bf e}{\bf e}^H]$ are diagonal. In
this case we can directly impose the effects of the uncertainties to
all the matrices associated with ${\bf f}_k$ and ${\bf g}$ in
\eqref{eq14}. By assuming that the channel errors are uncorrelated
with the channels so that $E[{\bf e}_k \odot {\bf g}]={\bf 0}$,
$E[{\bf e} \odot {\bf f}_k]={\bf 0}$, $E[{\bf e} \odot {\bf g}]={\bf
0}$ and $E[{\bf e}_k \odot {\bf f}_k]={\bf 0}$, then we can use an
additive Frobenius norm matrix perturbation
\cite{r8}, which results in
\begin{equation}
\hat{\bf R}_k={\bf R}_k+{\bf R}_{{\bf e}_k}={\bf R}_k+\epsilon||{\bf R}_k||_F{\bf I}_M, k=1,\dotsb,K, \label{eq16}
\end{equation}
\begin{equation}
\hat{\bf Q}={\bf Q}+{\bf R}_{\bf e}={\bf Q}+\epsilon||{\bf Q}||_F{\bf I}_M, k=1,\dotsb,K, \label{eq16+}
\end{equation}
\begin{equation}
\hat{\bf D}={\bf D}+\epsilon||{\bf D}||_F{\bf I}_M, \label{eq17}
\end{equation}
where $\hat{\bf R}_k$, $\hat{\bf Q}$ and $\hat{\bf D}$ are the
matrices perturbed after the channel mismatch effects are taken into
account, $\epsilon$ is the perturbation parameter uniformly
distributed within $(0,{\epsilon}_{max}]$ where ${\epsilon}_{max}$
is a predefined constant which describes the mismatch level. The
matrix ${\bf I}_M$ represents the identity matrix of dimension $M$
and it is clear that $\hat{\bf R}_k$, $\hat{\bf Q}$ and $\hat{\bf
D}$ are positive definite, i.e. $\hat{\bf R}_k \succ {\bf 0}
(k=1,\dotsb,K)$, $\hat{\bf Q} \succ {\bf 0}$ and $\hat{\bf D} \succ
{\bf 0}$.

RDB techniques compute the beamforming weights ${\bf w}$ such
that the output SINR can be maximized in the presence of
uncertainties in the channels. In particular, a robust design of
${\bf w}$ must solve the constrained optimization problem given by
\begin{equation}
\begin{aligned}
\underset{\bf w}{\rm max}~~ \frac{{\bf w}^H\hat{\bf R}_1{\bf w}}{P_n+{\bf w}^H(\hat{\bf Q}+\sum_{k=2}^K\hat{\bf R}_k){\bf w}} \\
{\rm subject} ~~ {\rm to} ~~~~ {\bf w}^H\hat{\bf D}{\bf w} \leq P_T.  \label{eq18}
\end{aligned}
\end{equation}
The optimization problem in \eqref{eq18} aims to maximize the output
SINR subject to a total relay transmit power constraint. Related
work has been discussed in \cite{r4}, where it has been shown that a
robust design of ${\bf w}$ can be computed in closed form using an
eigen-decomposition method that only requires quantities or
parameters with known second-order statistics. In this work, we aim
to cost-effectively solve \eqref{eq18} by exploiting low-rank
and cross-correlation techniques as described in what follows.

\section{Proposed LRCC-RDB Technique}

In this section, the LRCC-RDB technique is introduced and explained
in detail. The LRCC-RDB approach is suitable for systems with
imperfect CSI and is applicable to scenarios based on two different assumptions on CSI availability: \begin{itemize}
    \item when the instantaneous CSI is available  while  the  CSI second-order  statistics  is  not,  the channel covariance matrices are iteratively estimated and then the channel  error  spectrum  matrices are  iteratively  constructed.
    \item when the CSI second order statistics is available while the instantaneous CSI is not, the channel error spectrum matrices can be directly constructed at any iteration.
\end{itemize}

Here we detail LRCC-RDB following the former assumption, which
requires a few more steps due to the iterations and has a higher
complexity. We also note that the approach discussed in \cite{r4}
cannot be applied with this assumption. Therefore, the LRCC-RDB
technique works iteratively to estimate and obtain the channel
statistics over snapshots. The LRCC-RDB technique is based on the
exploitation of low-rank properties
\cite{intadap,inttvt,jio,ccmmwf,wlmwf,jidf,jidfecho,barc,jiols,jiomimo,jiostap,sjidf,l1stap,saabf,jioccm,ccmrab,wlbeam,lcrab,jiodoa,rrser,rcb,saalt,dce,damdc,locsme,memd,okspme,rrdoa,kaesprit,rhomo,sorsvd,corutv,sparsestap,wlccm,dcdrec,kacs}
and the cross-correction vector between the relay received data and
the system output. By projecting the so obtained cross-correlation
vector onto the subspace at the relays, the channel errors can be
efficiently mitigated and the result leads to a more precise
estimate of the mismatched channels. In the following exposition,
the snapshot index $i$ is introduced and the sample
cross-correlation vector (SCV) $\hat{\bf q}(i)$ associated with the
$i$th snapshot can be computed by
\begin{equation}
\hat{\bf q}(i)=\frac{1}{i}\sum\limits_{j=1}^i{\bf x}(j){z^*}(j), \label{eq19}
\end{equation}
which uses an averaging window that takes into account the snapshots
up to time index $i$, where ${\bf x}(i)$ and ${z^*}(i)$ refer to the
data observation vector in the $i$th snapshot at the relays and the
system output in the $i$th snapshot at the destination,
respectively, in the presence of channel uncertainties.

We then break down the mismatched channel matrix $\hat{\bf F}(i)$
into $K$ components as $\hat{\bf F}(i)=[\hat{\bf f}_1(i), \hat{\bf
f}_2(i), \dotsb,\hat{\bf f}_K(i)]$ and for each of them we construct
a separate projection matrix. The channel covariance matrices are estimated based on time-averaged sample matrices as
\begin{equation}
{\bf R}_{{\bf f}_k}(i)=\frac{1}{i}\sum\limits_{j=1}^i\hat{\bf f}_k(j)\hat{\bf f}_k^H(j), ~ \forall ~ k=1,\dotsb,K, \label{eq20}
\end{equation}
\begin{equation}
{\bf R}_{\bf g}(i)=\frac{1}{i}\sum\limits_{j=1}^i\hat{\bf g}(j)\hat{\bf g}^H(j). \label{eq20+}
\end{equation}

Then the error covariance matrices ${\bf R}_{{\bf e}_k}(i)$ and
${\bf R}_{\bf e}(i)$ can be computed as
\begin{equation}
{\bf R}_{{\bf e}_k}(i)=\epsilon||{\bf R}_{{\bf f}_k}(i)||_F{\bf I}_M. \label{eq21}
\end{equation}
\begin{equation}
{\bf R}_{{\bf e}}(i)=\epsilon||{\bf R}_{\bf g}(i)||_F{\bf I}_M. \label{eq21+}
\end{equation}
In order to reduce the impact of the errors ${\bf e}_k(i)$ from
$\hat{\bf f}_k(i)$ and ${\bf e}(i)$ from $\hat{\bf g}(i)$ on the
performance, the SCV obtained in \eqref{eq19} can be projected onto
the subspace as given by
\begin{multline}
{\bf P}_k(i)=[{\bf c}_{1,k}(i),{\bf c}_{2,k}(i),\dotsb,{\bf c}_{N,k}(i)] \\ [{\bf c}_{1,k}(i),{\bf c}_{2,k}(i),\dotsb,{\bf c}_{N,k}(i)]^H, \label{eq22}
\end{multline}
and
\begin{equation}
{\bf P}(i)=[{\bf c}_{1}(i),{\bf c}_{2}(i),\dotsb,{\bf c}_{N}(i)][{\bf c}_{1}(i),{\bf c}_{2}(i),\dotsb,{\bf c}_{N}(i)]^H, \label{eq22+}
\end{equation}
respectively, where ${\bf c}_{1,k}(i),{\bf c}_{2,k}(i),\dotsb,{\bf c}_{N,k}(i)$ and \\ ${\bf c}_{1}(i),{\bf c}_{2}(i),\dotsb,{\bf c}_{N}(i)$ are the $N$ principal eigenvectors of the error spectrum matrices ${\bf C}_k(i)$ and ${\bf C}(i)$ defined by
\begin{multline}
{\bf C}_k(i) \triangleq \int\limits_{\epsilon\rightarrow{0}^{+}}^{\epsilon_{max}}E[\hat{\bf f}_k(i)\hat{\bf f}_k^H(i)]d\epsilon \\
=\int\limits_{\epsilon\rightarrow{0}^{+}}^{\epsilon_{max}}E[({\bf f}_k(i)+{\bf e}_k(i))({\bf f}_k(i)+{\bf e}_k(i))^H]d\epsilon, \label{eq23} \\
\end{multline}
and
\begin{multline}
{\bf C}(i) \triangleq \int\limits_{\epsilon\rightarrow{0}^{+}}^{\epsilon_{max}}E[\hat{\bf g}(i)\hat{\bf g}^H(i)]d\epsilon \\
=\int\limits_{\epsilon\rightarrow{0}^{+}}^{\epsilon_{max}}E[({\bf g}(i)+{\bf e}(i))({\bf g}(i)+{\bf e}(i))^H]d\epsilon. \label{eq23+} \\
\end{multline}
respectively. The matrices ${\bf C}_k(i)$ and ${\bf C}(i)$ are low-rank matrices that can accurately represent the error components. Note that the selection of principal eigenvectors follows the following criterion: firstly, we select the eigenvectors corresponding to the eigenvalues larger than a manually set threshold determined by the noise level; secondly, we choose the eigenvectors whose eigenvalues are larger than the average value of all eigenvalues; thirdly, $N$ should be the minimum number that is also sufficiently large to retain most of the total variances of ${\bf C}_k(i) ~ \forall ~ k=1,\dotsb,K$ and ${\bf C}(i)$.


Since we have assumed that ${\bf e}_k(i)$ and ${\bf
e}(i)$ are uncorrelated with ${\bf f}_k(i)$ and ${\bf g}(i)$, if
$\epsilon$ follows a uniform distribution over the sector
$(0,\epsilon_{max}]$, by approximating $E[{\bf f}_k(i){\bf
f}_k^H(i)]\approx{\bf R}_{{\bf f}_k}(i)$, $E[{\bf e}_k(i){\bf
e}_k^H(i)]\approx{\bf R}_{{\bf e}_k}(i)$, $E[{\bf g}(i){\bf
g}^H(i)]\approx{\bf R}_{\bf g}(i)$ and $E[{\bf e}(i){\bf
e}^H(i)]\approx{\bf R}_{\bf e}(i)$ based on the sample covariance matrices in \eqref{eq20} and \eqref{eq20+}, \eqref{eq23} and \eqref{eq23+}
can be simplified to
\begin{multline}
{\bf C}_k(i)=\int\limits_{\epsilon\rightarrow{0}^{+}}^{\epsilon_{max}}({\bf R}_{{\bf f}_k}(i)+{\bf R}_{{\bf e}_k}(i))d\epsilon \\
=\epsilon_{max}{\bf R}_{{\bf f}_k}(i)+\frac{\epsilon_{max}^2}{2}||{\bf R}_{{\bf f}_k}(i)||_F{\bf I}_M, \label{eq24}
\end{multline}
and
\begin{multline}
{\bf C}(i)=\int\limits_{\epsilon\rightarrow{0}^{+}}^{\epsilon_{max}}({\bf R}_{\bf g}(i)+{\bf R}_{\bf e}(i))d\epsilon \\
=\epsilon_{max}{\bf R}_{\bf g}(i)+\frac{\epsilon_{max}^2}{2}||{\bf R}_{\bf g}(i)||_F{\bf I}_M. \label{eq24+}
\end{multline}
We remark that when the system only knows the second-order statistics of the CSI instead of the instantaneous CSI, \eqref{eq20} and \eqref{eq20+} are not required, and the steps from \eqref{eq21} to \eqref{eq24+} are performed without iterations, which means they are only performed once to obtain ${\bf C}_k(i) ~ \forall ~ k,\dotsb, K$ and ${\bf C}(i)$.

\begin{figure}[!htb]
 \begin{center}
 \includegraphics[width=0.95\columnwidth]{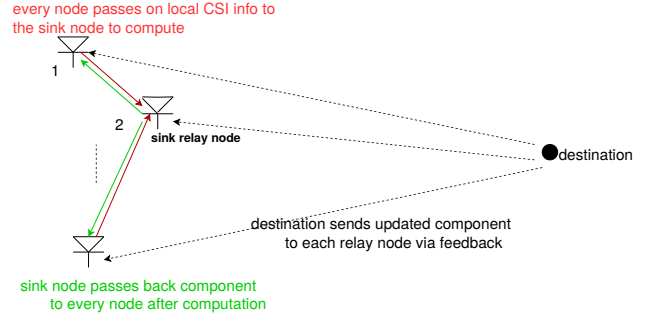}
 \caption{Distributed beamforming scheme based on CSI availability}
 \label{distributed}
 \end{center}
\end{figure}

Next, by projecting $\hat{\bf q}(i)$ onto ${\bf P}_k(i)$ and ${\bf P}(i)$, we can eliminate uncorrelated information in the orthogonal subspace and only extract the correlated information (i.e. the channel estimates $\hat{\bf f}_k(i)$ and $\hat{\bf g}(i)$, respectively) that exist in both. The mismatched channel components are then estimated by
\begin{equation}
\hat{\bf f}_k(i)=\frac{{\bf P}_k(i)\hat{\bf q}(i)}{{\lVert{{\bf P}_k(i)\hat{\bf q}(i)}\rVert}_2}, \label{eq25}
\end{equation}
\begin{equation}
\hat{\bf g}(i)=\frac{{\bf P}(i)\hat{\bf q}(i)}{{\lVert{{\bf P}(i)\hat{\bf q}(i)}\rVert}_2}. \label{eq25+}
\end{equation}
To this point, all the $K$ channel components of $\hat{\bf f}_k(i)$
can be obtained so that we have $\hat{\bf F}_k(i)=[\hat{\bf f}_1(i),
\hat{\bf f}_2(i), \dotsb, \hat{\bf f}_K(i)]$. In the next step, we
will use the estimated mismatched channel components to provide
estimates for the matrix quantities $\hat{\bf R}_k(i)$
($k=1,\dotsb,K$), $\hat{\bf Q}(i)$ and $\hat{\bf D}(i)$ in
\eqref{eq18} as follows:
\begin{equation}
\hat{\bf R}_k(i)=P_{s,k}E[(\hat{\bf f}_k(i) \odot \hat{\bf g}(i))(\hat{\bf f}_k(i) \odot \hat{\bf g}(i))^H], \label{eq26}
\end{equation}
\begin{equation}
\hat{\bf Q}(i)=P_nE[\hat{\bf g}(i)\hat{\bf g}^H(i)], \label{eq26+}
\end{equation}
\begin{equation}
\hat{\bf D}(i)={\rm diag}\Big(\sum_{k=1}^KP_{s,k}[E[|\hat{f}_{1,k}(i)|^2], \dotsb, E[\hat{f}_{M,k}(i)|^2]]+P_n\Big). \label{eq27}
\end{equation}
To proceed further, we define $\hat{\bf U}(i)=\hat{\bf Q}(i)+\sum_{k=2}^K\hat{\bf R}_k(i)$ so that \eqref{eq18} can be written as
\begin{equation}
\begin{aligned}
\underset{{\bf w}(i)}{\rm max}~~ \frac{{\bf w}^H(i)\hat{\bf R}_1(i){\bf w}(i)}{P_n+{\bf w}^H(i)\hat{\bf U}(i){\bf w}(i)} \\
{\rm subject} ~~ {\rm to} ~~~~ {\bf w}^H(i)\hat{\bf D}(i){\bf w}(i) \leq P_T.  \label{eq28} \\
\end{aligned}
\end{equation}
The solution of the optimization problem in \eqref{eq28} is given by
the weight vector
\begin{equation}
{\bf w}(i)=\sqrt{p}{\bf D}^{-1/2}(i)\tilde{\bf w}(i), \label{eq29}
\end{equation}
where $\tilde{\bf w}(i)$ satisfies $\tilde{\bf w}^H(i)\tilde{\bf
w}(i)=1$. Then \eqref{eq28} can be rewritten as
\begin{equation}
\begin{aligned}
\underset{p,\tilde{\bf w}(i)}{\rm max}~~\frac{p\tilde{\bf w}^H(i)\tilde{\bf R}_1(i)\tilde{\bf w}(i)}{p\tilde{\bf w}^H(i)\tilde{\bf U}(i)\tilde{\bf w}(i)+P_n} \\
{\rm subject} ~~ {\rm to} ~~~~ ||\tilde{\bf w}(i)||^2=1, p \leq P_T, \label{eq30}
\end{aligned}
\end{equation}
where $\tilde{\bf R}_1(i)=\hat{\bf D}^{-1/2}(i)\hat{\bf R}_1(i){\bf
D}^{-1/2}(i)$ and $\tilde{\bf U}(i)=\hat{\bf D}^{-1/2}(i)\hat{\bf U}(i)\hat{\bf
D}^{-1/2}(i)$. As the objective function in \eqref{eq30} increases
monotonically with $p$ regardless of $\tilde{\bf w}(i)$, which means
the objective function is maximized when $p=P_T$, hence \eqref{eq30}
can be simplified to
\begin{equation}
\begin{aligned}
\underset{\tilde{\bf w}(i)}{\rm max}~~\frac{P_T\tilde{\bf w}^H(i)\tilde{\bf R}_1(i)\tilde{\bf w}(i)}{P_T\tilde{\bf w}^H(i)\tilde{\bf U}(i)\tilde{\bf w}(i)+P_n} \\
{\rm subject} ~~ {\rm to} ~~~~ ||\tilde{\bf w}(i)||^2=1, \label{eq31}
\end{aligned}
\end{equation}
or equivalently as
\begin{equation}
\begin{aligned}
\underset{\tilde{\bf w}(i)}{\rm max}~~\frac{P_T\tilde{\bf w}^H(i)\tilde{\bf R}_1(i)\tilde{\bf w}(i)}{\tilde{\bf w}^H(i)(P_n{\bf I}_M+P_T\tilde{\bf U}(i))\tilde{\bf w}(i)} \\
{\rm subject} ~~ {\rm to} ~~~~ ||\tilde{\bf w}(i)||^2=1, \label{eq32}
\end{aligned}
\end{equation}
in which the objective function is maximized when $\tilde{\bf w}(i)$
is chosen as the principal eigenvector of $(P_n{\bf
I}_M+P_T{\tilde{\bf U}(i)})^{-1}\tilde{\bf R}_1(i)$ \cite{r4}, which
leads to the solution for the weight vector of the LRCC-RDB
technique given by
\begin{multline}
{\bf w}(i)=\sqrt{P_T}\hat{\bf D}^{-1/2}(i){\mathcal P}\{(P_n{\bf I}_M \\ +\hat{\bf D}^{-1/2}(i)\hat{\bf U}(i)\hat{\bf D}^{-1/2}(i))^{-1} \hat{\bf D}^{-1/2}(i)\hat{\bf R}_1(i)\hat{\bf D}^{-1/2}(i)\}, \label{eq33}
\end{multline}
where ${\mathcal P}\{.\}$ denotes the principal eigenvector
corresponding to the largest eigenvalue. As already pointed out, there is no online optimization required for the proposed LRCC-RDB technique. The information required to be shared among the relay nodes includes the destination signal $z(i)$ which is a scalar and can be fed back from the destination; the signals observed at each relay node $x_m(i)$ ($m=1,\dotsb, M$) that are only exchanged among the relays; and the system CSI which is used to compute the optimized weight vector ${\bf w}(i)$ in the given close-form expression \eqref{eq33}. However, the devise in which this computation must be done depends on where the system CSI is available to avoid signaling overhead. For example, if the CSI is available at the relay nodes, then it is best to compute ${\bf w}(i)$ at the relay nodes which can be accomplished by setting a sink node to perform the matrix computations, then each component (a scalar) of the computed ${\bf w}(i)$ is sent to the corresponding relay node for update via node cooperation; if the CSI is available at the destination, then the computation of ${\bf w}(i)$ is carried out at the destination before each component (a scalar) of the computed ${\bf w}(i)$ is sent back to the corresponding relay node for update via feedback, as explained in Fig. \ref{distributed}. Note that the information exchange of scalars among the relays is not necessary for every snapshot in scenarios where the channel is static for multiple snapshots. In this case, the information exchange only needs to be performed once per data block comprising multiple snapshots instead of every snapshot.

Then the maximum achievable
SINR of the system in the presence of channel errors is given by
\begin{multline}
{\rm SINR}_{max}=P_T{\lambda}_{\rm largest}\{(P_n{\bf I}_M+\hat{\bf D}^{-1/2}(i)\hat{\bf U}(i)\hat{\bf D}^{-1/2}(i))^{-1} \\ \hat{\bf D}^{-1/2}(i)\hat{\bf R}_1(i)\hat{\bf D}^{-1/2}(i)\}, \label{eq34}
\end{multline}
where the operator $\lambda_{\rm largest}\{.\}$ extracts the largest eigenvalue of the argument. The steps of the
proposed LRCC-RDB technique are detailed in Table \ref{table1}.

\begin{table}
\small
\begin{center}
\caption{Proposed LRCC-RDB Technique}
\begin{tabular}{l}
\hline
Initialization: \\
${\bf w}(0)={\bf 1}$; $\hat{\bf q}(0)={\bf 1}$; ${\bf R}_{{\bf f}_k}(0)=0.01{\bf I}_M$ for $k=1,\dotsb,K$; \\
${\bf R}_{\bf g}(0)=0.01{\bf I}_M$; $\epsilon_{max}$; $N$; $P_T$;
{$\epsilon_{max}$}.
\\\\
For iteration $i=1,2,\dotsb$: \\
~~~~Compute the SCV as:  \\
~~~~$\hat{\bf q}(i)=((i-1)\cdot\hat{\bf q}(i-1)+{\bf
x}(i){z^*}(i))/i$ \\\\
~~~~For $k=1,\dotsb,K$:   \\
~~~~~~~~Estimate the channel covariance matrices:  \\
~~~~~~~~${\bf R}_{{\bf f}_k}(i)\approx((i-1)\cdot{\bf R}_{{\bf f}_k}(i-1)+\hat{\bf f}_k(i)\hat{\bf f}_k^H(i))/i$ \\
~~~~~~~~${\bf R}_{\bf g}(i)\approx((i-1)\cdot{\bf R}_{{\bf g}}(i-1)+\hat{\bf g}(i)\hat{\bf g}^H(i))/i$
\\\\
~~~~~~~~Compute the error spectrum matrices for $\hat{\bf f}_k(i)$ and $\hat{\bf g}(i)$:  \\
~~~~~~~~${\bf C}_k(i)=\epsilon_{max}{\bf R}_{{\bf f}_k}(i)+\frac{\epsilon_{max}^2}{2}||{\bf R}_{{\bf f}_k}(i)||_F{\bf I}_M$ \\
~~~~~~~~${\bf C}(i)=\epsilon_{max}{\bf R}_{\bf g}(i)+\frac{\epsilon_{max}^2}{2}||{\bf R}_{\bf g}(i)||_F{\bf I}_M$
\\\\
~~~~~~~~Compute $N$ principal eigenvectors of ${\bf C}_k(i)$ and ${\bf C}(i)$: \\
~~~~~~~~$[{\bf c}_{1,k}(i),{\bf c}_{2,k}(i),\dotsb,{\bf c}_{N,k}(i)]$ and $[{\bf c}_{1}(i),{\bf c}_{2}(i),\dotsb,{\bf c}_{N}(i)]$
\\\\
~~~~~~~~Compute the projection matrix for $\hat{\bf f}_k(i)$ and $\hat{\bf g}(i)$:  \\
~~~~~~~~${\bf P}_k(i)=[{\bf c}_{1,k}(i),{\bf c}_{2,k}(i),\dotsb,{\bf c}_{N,k}(i)]$ \\
~~~~~~~~$[{\bf c}_{1,k}(i),{\bf c}_{2,k}(i),\dotsb,{\bf c}_{N,k}(i)]^H$ \\
~~~~~~~~${\bf P}(i)=[{\bf c}_{1}(i),{\bf c}_{2}(i),\dotsb,{\bf c}_{N}(i)][{\bf c}_{1}(i),{\bf c}_{2}(i),\dotsb,{\bf c}_{N}(i)]^H$
\\\\
~~~~~~~~Estimate $\hat{\bf f}_k(i)$ and $\hat{\bf g}(i)$ by subspace projections:  \\
~~~~~~~~$\hat{\bf f}_k(i)=\frac{{\bf P}_k(i)\hat{\bf q}(i)}{{\lVert{{\bf P}_k(i)\hat{\bf q}(i)}\rVert}_2}$ \\
~~~~~~~~$\hat{\bf g}(i)=\frac{{\bf P}(i)\hat{\bf q}(i)}{{\lVert{{\bf P}(i)\hat{\bf q}(i)}\rVert}_2}$
\\\\
~~~~~~~~Compute $\hat{\bf R}_k(i)$:  \\
~~~~~~~~$\hat{\bf R}_k(i)=P_{s,k}E[(\hat{\bf f}_k(i) \odot \hat{\bf g}(i))(\hat{\bf f}_k(i) \odot \hat{\bf g}(i))^H]$
\\
~~~~End of $k$. \\\\
~~~~Compute quantities $\hat{\bf D}(i)$, $\hat{\bf Q}(i)$ and $\hat{\bf U}(i)$: \\
~~~~$\hat{\bf D}(i)={\rm diag}(\sum_{k=1}^KP_{s,k}[E[|\hat{f}_{1,k}(i)|^2], E[|\hat{f}_{2,k}(i)|^2], \dotsb,$ \\ ~~~~$E[\hat{f}_{M,k}(i)|^2]]+P_n)$
\\
~~~~$\hat{\bf Q}(i)=P_nE[\hat{\bf g}(i)\hat{\bf g}^H(i)]$ \\
~~~~$\hat{\bf U}(i)=\hat{\bf Q}(i)+\sum_{k=2}^K\hat{\bf R}_k(i)$ \\\\
~~~~Obtain the LRCC-RDB weight vector: \\
~~~~${\bf w}(i)=\sqrt{P_T}\hat{\bf D}^{-1/2}(i){\mathcal P}\{(P_n{\bf I}_M+\hat{\bf D}^{-1/2}(i)\hat{\bf U}(i)\hat{\bf D}^{-1/2}(i))^{-1}$
\\
~~~~$\hat{\bf D}^{-1/2}(i)\hat{\bf R}_1(i)\hat{\bf D}^{-1/2}(i)\}$
\\\\
~~~~Compute the system output SINR: \\
~~~~${\rm SINR}_{max}=P_T{\lambda}_{\rm largest}\{(P_n{\bf I}_M+\hat{\bf
D}^{-1/2}(i)\hat{\bf U}(i)\hat{\bf D}^{-1/2}(i))^{-1}$  \\
~~~~$\hat{\bf D}^{-1/2}(i)\hat{\bf R}_1(i)\hat{\bf D}^{-1/2}(i)\}$
\\
End of $i$. \\
\hline
\end{tabular} \label{table1}
\end{center}
\end{table}

\section{Analysis}

This section presents a performance analysis of the proposed LRCC-RDB technique in terms of the MSEs for the channels and its complexity. Although the performance analysis is for distributed beamforming the main principles can also be useful for other scenarios and techniques. In the MSE analysis, we emphasize the required assumptions that the channel components ${\bf f}_k$,
$k=1,\dotsb,K$, ${\bf g}$, the error vectors ${\bf e}_k$,
$k=1,\dotsb,K$, ${\bf e}$ and the noise ${\bf\nu}$, $n$ are all
uncorrelated with each other. We investigate the MSE performance using two different approaches, one obtains a pair of upper and lower bounds that are based on the spread of the channel covariance matrix for the channel error model adopted, which are useful to assess most RDB techniques, whereas the other approach obtains tighter bounds for subspace projection-based techniques such as the proposed LRCC-RDB technique that involves the SCV and are related to principal component analysis (PCA).

\subsection{{MSE Analysis}}


In this section, we carry out a general MSE analysis of the channel
errors associated with the distributed beamforming problem. The
objectives of the proposed MSE analysis are to provide an analytic
investigation of the proposed LRCC-RDB and existing techniques and
establish the following:
\begin{itemize}
  \item to obtain a pair of upper and lower bounds for RDB methods that model the channel error covariance
      matrix as additive perturbation based on the Frobenius norm of the true channel covariance matrix
      and a perturbation parameter $\epsilon$ as in \eqref{eq21}.

  \item to show that the proposed LRCC-RDB algorithm outperforms the RDB methods, which generally employ
  an additive perturbation for the channel mismatches and do not exploit
  prior knowledge in the form of cross-correlation and subspace structures like LRCC-RDB.
\end{itemize}

Let us first define the MSE between ${\bf f}_k$ and $\hat{\bf f}_k$
as
\begin{equation}
\begin{split}
{\rm MSE}\{\hat{\bf f}_k\}_1 & \triangleq {\rm tr}(E[(\hat{\bf f}_k-{\bf f}_k)(\hat{\bf f}_k-{\bf f}_k)^H])  \\
& ={\rm tr}(E[{\bf e}_k{\bf e}_k^H]) ={\rm tr}({\bf R}_{{\bf e}_k})  \\
& ={\rm tr} \Big(\frac{{\epsilon}_{max}}{2}||{\bf R}_{{\bf f}_k}||_F{\bf
I}_M \Big)=\frac{{\epsilon}_{max}M}{2}||{\bf R}_{{\bf f}_k}||_F.
\label{eq35}
\end{split}
\end{equation}
Furthermore, the Frobenius norm of any positive definite matrix can
be expressed as the square root of the sum of its squared
eigenvalues, which results in
\begin{equation}
||{\bf R}_{{\bf f}_k}||_F=\sqrt{\sum_{m=1}^M\lambda_{m,k}^2}, \label{eq36}
\end{equation}
where $\lambda_{m,k}$ refers to the $m$th eigenvalue of matrix ${\bf R}_{{\bf f}_k}$.

\begin{figure}[!htb]
\begin{center}
\def\epsfsize#1#2{0.99\columnwidth}
\epsfbox{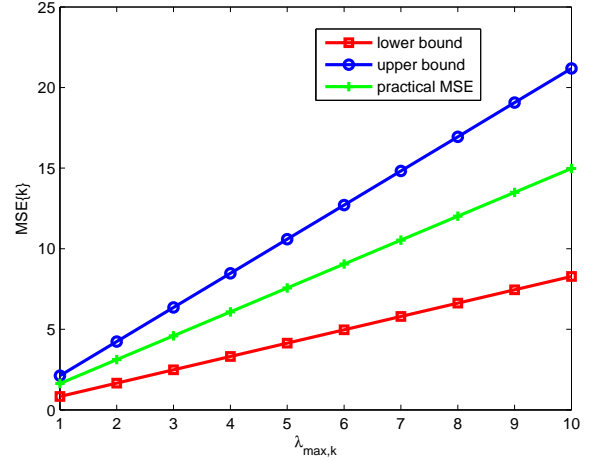}
\caption{MSE bounds versus $\lambda_{max,k}$, $\sigma_{\lambda,k}=0.9\lambda_{max,k}$} \label{figure09}
\end{center}
\end{figure}

\begin{figure}[!htb]
\begin{center}
\def\epsfsize#1#2{0.99\columnwidth}
\epsfbox{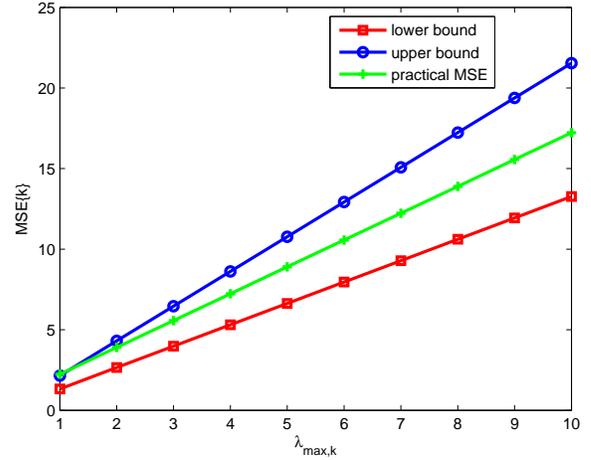}
\caption{MSE bounds versus $\lambda_{max,k}$, $\sigma_{\lambda,k}=0.5\lambda_{max,k}$} \label{figure05}
\end{center}
\end{figure}

Let us now denote the eigenvalue spread of the matrix ${\bf R}_{{\bf
f}_k}$ as $\sigma_{\lambda,k}$, which is defined by
$|\lambda_{max,k}-\lambda_{min,k}|$, where $\lambda_{max,k}$ and
$\lambda_{max,k}$ refer to the maximum eigenvalue and the minimum
eigenvalue of ${\bf R}_{{\bf f}_k}$, respectively. Then we can
obtain a lower bound for ${\rm min}\{\sum_{m=1}^M\lambda_{m,k}^2\}$
by assuming $\lambda_{1,k}, \lambda_{2,k}, \dotsb$, $\lambda_{m,k},
\dotsb, \lambda_{M,k}$ ($\lambda_{m,k}\neq\lambda_{max,k}$)
$\rightarrow 0^+$, which yields the following relations for the
lower bound on the MSE of $\hat{\bf f}_k$:
\begin{equation}
\begin{split}
{\rm min}\Big\{\sum_{m=1}^M\lambda_{m,k}^2\Big\} & > (M-1)\lambda_{min,k}^2+\lambda_{max,k}^2  \\
&= (M-1) (\lambda_{max,k}-\sigma_{\lambda,k})^2+\lambda_{max,k}^2  \\
&= M \lambda_{max,k}^2-2(M-1)\sigma_{\lambda,k}\lambda_{max,k} \\
&+(M-1)\sigma_{\lambda,k}^2.
\label{eq37}
\end{split}
\end{equation}
We can also obtain an upper bound for ${\rm
max}\{\sum_{m=1}^M\lambda_{m,k}^2\}$ by assuming $\lambda_{1,k},
\lambda_{2,k}, \dotsb$, $\lambda_{m,k}, \dotsb, \lambda_{M,k}$
($\lambda_{m,k}\neq\lambda_{max,k}$) $\rightarrow
\lambda_{max,k}^-$, which yields the following relations for the
upper bound on the MSE of $\hat{\bf f}_k$:
\begin{equation}
\begin{split}
{\rm max}\Big\{\sum_{m=1}^M\lambda_{m,k}^2\Big\} & < (M-1)\lambda_{max,k}^2+\lambda_{min,k}^2  \\
& = (M-1)\lambda_{max,k}^2+(\lambda_{max,k}-\sigma_{\lambda,k})^2  \\
& = M \lambda_{max,k}^2 -2\sigma_{\lambda,k}
\lambda_{max,k}+\sigma_{\lambda,k}^2. \label{eq38}
\end{split}
\end{equation}
By substituting \eqref{eq36} into \eqref{eq37} and \eqref{eq38}, we
obtain
\begin{multline}
{\rm min}\{||{\bf R}_{{\bf f}_k}||_F\} > \\ \sqrt{M\lambda_{max,k}^2-2(M-1)\sigma_{\lambda,k}\lambda_{max,k}+(M-1)\sigma_{\lambda,k}^2}, \label{eq39}
\end{multline}
\begin{multline}
{\rm max}\{||{\bf R}_{{\bf f}_k}||_F\} < \sqrt{M\lambda_{max,k}^2-2\sigma_{\lambda,k}\lambda_{max,k}+\sigma_{\lambda,k}^2}. \label{eq40}
\end{multline}
Since we have ${\rm min}\{||{\bf R}_{{\bf f}_k}||_F\} \leq ||{\bf
R}_{{\bf f}_k}||_F \leq {\rm max}\{||{\bf R}_{{\bf f}_k}||_F\},$
then we can obtain the upper and lower bounds for $||{\bf R}_{{\bf
f}_k}||_F$ by substituting the relations in \eqref{eq37} and
\eqref{eq38} in \eqref{eq39} and \eqref{eq40}, respectively, resulting in
\begin{multline}
\sqrt{M\lambda_{max,k}^2-2(M-1)\sigma_{\lambda,k}\lambda_{max,k}+(M-1)\sigma_{\lambda,k}^2} \\
< ||{\bf R}_{{\bf f}_k}||_F < \\
\sqrt{M\lambda_{max,k}^2-2\sigma_{\lambda,k}\lambda_{max,k}+\sigma_{\lambda,k}^2}, \label{eq41}
\end{multline}
which is then substituted in \eqref{eq35} and yields the bounds for the MSE:
\begin{multline}
\frac{{\epsilon}_{max}M}{2}\sqrt{M\lambda_{max,k}^2-2(M-1)\sigma_{\lambda,k}\lambda_{max,k}+(M-1)\sigma_{\lambda,k}^2} \\
< {\rm MSE}\{\hat{\bf f}_k\}_1 < \\
\frac{{\epsilon}_{max}M}{2}\sqrt{M\lambda_{max,k}^2-2\sigma_{\lambda,k}\lambda_{max,k}+\sigma_{\lambda,k}^2}. \label{eq42}
\end{multline}
The bounds described in \eqref{eq42} give an insight on how the MSE
of the $k$th component of $\hat{\bf F}$ is bounded by the maximum
(principal) eigenvalue $\lambda_{max,k}$ of the channel covariance
matrix ${\bf R}_{{\bf f}_k}$ and the eigenvalue spread
$\sigma_{\lambda,k}$ of its Frobenius norm $||{\bf R}_{{\bf
f}_k}||_F$. The same procedure can be carried out for analyzing the
MSE of channel $\hat{\bf g}$, which is not presented here to avoid a
repetitive development. We can employ both lower bounds of the channel components $\hat{\bf f}_k$ and channel $\hat{\bf g}$ as their minimum
MSEs (MMSEs) to compute the MMSEs of $\hat{\bf F}$ and $\hat{\bf g}$ as
\begin{multline}
{\rm MMSE}\{\hat{\bf F}\}=\sum_{k=1}^K{\rm MMSE}\{\hat{\bf f}_k\}
=\frac{{\epsilon}_{max}M}{2} \\ \sum_{k=1}^K\sqrt{M\lambda_{max,k}^2-2(M-1)\sigma_{\lambda,k}\lambda_{max,k}+(M-1)\sigma_{\lambda,k}^2},
\label{eqa}
\end{multline}
\begin{equation}
\begin{split}
    & {\rm MMSE}\{\hat{\bf g}\}=\frac{{\epsilon}_{max}M}{2} \\ & \sqrt{M\lambda_{max}^2-2(M-1)\sigma_{\lambda}\lambda_{max}+(M-1)\sigma_{\lambda}^2}, \label{eqa+}
\end{split}
\end{equation}
respectively. Note that $\lambda_{max}$ here is a scalar representing the largest eigenvalue of ${\bf R}_{\bf g}$ and $\sigma_{\lambda}$ is the corresponding eigenvalue spread.
As an example, we
set the total number of relays and signal sources $M=8$,
$\epsilon_{max}=0.2$ and the input SNR is set to $10$dB. Then we
test two cases with $\sigma_{\lambda,k}=0.9\lambda_{max,k}$ and
$\sigma_{\lambda,k}=0.5\lambda_{max,k}$ and illustrate the
variations of those bounds in Figs. \ref{figure09} and
\ref{figure05}, respectively. Because we use linear relations
between $\sigma_{\lambda,k}$ and $\lambda_{max,k}$, proportional
relations between the MSE bounds and $\lambda_{max,k}$ are reflected
as can be seen in Figs. \ref{figure09} and \ref{figure05}. In
addition, we generate the signals that are processed by the sensor
array and compute the actual MSE values of the mismatched channels which are independent of the RDB algorithms, according to the above conditions and compare the results to
the analytic bounds in Figs. \ref{figure09} and \ref{figure05}.
The results are obtained by
taking the average MSE result from $k=1,\dotsb,K$. 
The sets of matrix eigenvalues are selected to be as close as
possible to the analytic conditions assumed for the sake of
comparison. Also, by comparing the values and variations of the MSE
bounds in Figs. \ref{figure09} and \ref{figure05}, we can see that
there is no obvious difference between the upper bounds. However,
with a smaller eigenvalue spread $\sigma_{\lambda,k}$, the lower
bound gets closer to the upper bound. The results obtained by
generating the signals processed by the sensor array indicate that
with a small $\lambda_{max,k}$ the MSE gets closer to the upper
bound.

\subsection{Analysis of Low-rank and Cross-Correlation Processing}

In this section, we present the performance analysis of the proposed
LRCC-RDB technique. In particular, this analysis is specific to the
low-rank and cross-correlation processing performed by the proposed
LRCC-RDB method. At first we aim to exploit the properties of the
cross-correlation vector ${\bf q}(i)$ estimated in \eqref{eq19}. For
convenience, we omit the time index $i$ in the following
analysis. By definition, we have
\begin{equation}
\begin{split}
{\bf q}  \triangleq E[z^*{\bf x}]& =E[(\hat{\bf g}^T{\bf W}{\bf x}+n)^*{\bf x}] \\
& =E[(\hat{\bf g}^H{\bf W}^*{\bf x}^*+n^*)(\hat{\bf F}{\bf s}+{\bf
\nu})]. \label{eq43}
\end{split}
\end{equation}
Since ${\bf W}$ is diagonal, we have ${\bf W}^*={\bf W}^H$. By
assuming that the noise $n$ is uncorrelated with ${\bf s}$ and ${\bf
\nu}$, the terms $E[n^*\hat{\bf F}{\bf s}]$ and $E[n^*{\bf \nu}]$
are equal to zero and can be discarded. Then from \eqref{eq43} we
have
\begin{equation}
\begin{split}
{\bf q}& =E[\hat{\bf g}^H{\bf W}^*{\bf x}^*\hat{\bf F}{\bf s}+\hat{\bf g}^H{\bf W}^*{\bf x}^*{\bf \nu}] \\
&=E[\hat{\bf g}^H{\bf W}^*(\hat{\bf F}^*{\bf s}^*+{\bf \nu}^*)\hat{\bf F}{\bf s}+\hat{\bf g}^H{\bf W}^*{\bf x}^*{\bf \nu}] \\
&=E[\hat{\bf F}{\bf s}{\bf s}^H\hat{\bf F}^H{\bf W}\hat{\bf
g}]+E[{\bf \nu}{\bf \nu}^H{\bf W}\hat{\bf g}], \label{eq44}
\end{split}
\end{equation}
where $\hat{\bf F}=[\hat{\bf f}_1,\dotsb,\hat{\bf f}_K]$ is the
mismatched channel matrix and ${\bf s}=[s_1,\dotsb,s_K]^T$. Now we
expand the expressions for both $\hat{\bf F}$ and ${\bf s}$ in
\eqref{eq44} while assuming that the source signals are uncorrelated
with each other. Then we obtain
\begin{equation}
\begin{split}
{\bf q}& =E[(\sum_{k=1}^K\hat{\bf f}_k{s}_k)(\sum_{k=1}^K\hat{\bf f}_k{s}_k)^H{\bf W}\hat{\bf g}]+E[{\bf \nu}{\bf \nu}^H{\bf W}\hat{\bf g}] \\
& =E[\sum_{k=1}^K{s}_k{s}_k^*\hat{\bf f}_k\hat{\bf f}_k^H{\bf
W}\hat{\bf g}]+E[{\bf \nu}{\bf \nu}^H{\bf W}\hat{\bf g}].
\label{eq45}
\end{split}
\end{equation}
At this stage, we emphasize that the analysis procedure for channel
$\hat{\bf f}_k$ is independent from $\hat{\bf g}$ as we will project
the same cross-correlation vector onto their subspace independently
even though the procedures are the same. For ease of exposition and
comparison with the MSE analysis of the previous subsection, we
focus on the analysis for ${\bf f}_k$ only and employ
${\bf\xi}=E[{\bf W}\hat{\bf g}]$. Therefore, if we substitute
$\hat{\bf f}_k={\bf f}_k+{\bf e}_k$ in \eqref{eq45} and replace
$E[{s}_k{s}_k^*]$ and $E[{\bf \nu}{\bf \nu}^H]$ with $P_{s,k}$ and
$P_n$, respectively, \eqref{eq45} can be simplified to
\begin{equation}
\begin{split}
{\bf q} & =E[\sum_{k=1}^KP_{s,k}{\bf f}_k{\bf f}_k^H+{\bf e}_k{\bf e}_k^H]{\bf\xi}+P_n{\bf\xi} \\
& =(\sum_{k=1}^KP_{s,k}E[({\bf R}_{{\bf f}_k}+{\bf R}_{{\bf
e}_k})]+P_n){\bf\xi}. \label{eq46}
\end{split}
\end{equation}
Let us now define the $k$th cross-correlation vector component as
\begin{equation}
{\bf q}_k \triangleq (P_{s,k}E[({\bf R}_{{\bf f}_k}+{\bf R}_{{\bf e}_k})]){\bf\xi}, \label{eq47}
\end{equation}
and write the cross-correlation vector as
\begin{equation}
{\bf q} \triangleq \sum_{k=1}^K{\bf q}_k+P_n{\bf\xi}. \label{eq48}
\end{equation}
LRCC-RBD applies a subspace projection to the
cross-correlation vector by substituting \eqref{eq48} in $\hat{\bf
f}_k={\bf P}_k{\bf q}$ (assuming it is normalized as in
\eqref{eq25}), which results in
\begin{equation}
\hat{\bf f}_k={\bf P}_k(\sum_{k=1}^K{\bf q}_k+P_n{\bf\xi}). \label{eq49}
\end{equation}
If we assume that by projecting any cross-correlation vector
component ${\bf q}_l$ generated from the channel components
$\hat{\bf f}_l$ ($1\leq l (l\neq k) \leq K$) onto the subspace
projection matrix ${\bf P}_k$ we have ${\bf P}_k{\bf q}_l=0$, then
\eqref{eq49} can be simplified to
\begin{equation}
\hat{\bf f}_k  ={\bf P}_k({\bf q}_k+P_n{\bf\xi})
={\bf P}_k(P_{s,k}E[({\bf R}_{{\bf f}_k}+{\bf R}_{{\bf
e}_k})]+P_n){\bf\xi}. \label{eq50}
\end{equation}
From the MSE definition in \eqref{eq35}, we have $${\rm
MSE}\{\hat{\bf f}_k\}_2={\rm tr}(E[(\hat{\bf f}_k-{\bf
f}_k)(\hat{\bf f}_k-{\bf f}_k)^H])=E[(\hat{\bf f}_k-{\bf
f}_k)^H(\hat{\bf f}_k-{\bf f}_k)].$$

After substituting \eqref{eq50} in \eqref{eq35}, we obtain
\begin{equation}
\begin{split}
{\rm MSE}\{\hat{\bf f}_k\}_2 & =E[({\bf P}_k(P_{s,k}E[({\bf R}_{{\bf f}_k}+{\bf R}_{{\bf e}_k})]+P_n){\bf\xi}-{\bf f}_k)^H  \\
& \quad ({\bf P}_k(P_{s,k}E[({\bf R}_{{\bf f}_k}+{\bf R}_{{\bf
e}_k})]+P_n){\bf\xi}-{\bf f}_k)].   \label{eq51}
\end{split}
\end{equation}
After expanding \eqref{eq51}, we get
\begin{equation}
\begin{split}
{\rm MSE}\{\hat{\bf f}_k\}_2 & =E[{\bf\xi}^H(P_{s,k}{\bf R}_{{\bf f}_k}+P_{s,k}{\bf R}_{{\bf e}_k}+P_n){\bf P}_k^H{\bf P}_k(P_{s,k}  \\
& \quad {\bf R}_{{\bf f}_k}+P_{s,k}{\bf R}_{{\bf e}_k}+P_n){\bf\xi}]  -2E[{\bf f}_k^H{\bf P}_k(P_{s,k}{\bf R}_{{\bf f}_k}  \\
& \quad +P_{s,k}{\bf R}_{{\bf e}_k}+P_n){\bf\xi}] +E[{\bf f}_k^H{\bf
f}_k]. \label{eq52}
\end{split}
\end{equation}
It should be noticed that ${\bf P}_k^H{\bf P}_k={\bf P}_k={\bf
P}_k^H$ as the projection of a subspace projection matrix onto
itself results in the same projection matrix. In addition, we have
${\bf f}_k^H{\bf P}_k={\bf f}_k^H$ in the second term of
\eqref{eq52}, which can be verified as follows. Since we have ${\bf
f}_k=E[{\bf P}_k{\bf q}_k]$ by pre-multiplying both sides by ${\bf
P}_k^H$ then we have ${\bf P}_k^H{\bf f}_k=E[{\bf P}_k^H{\bf
P}_k{\bf q}_k]=E[{\bf P}_k{\bf q}_k]={\bf f}_k$. Then by taking the
Hermitian transpose on both sides gives ${\bf f}_k^H{\bf P}_k={\bf
f}_k^H$. Therefore, \eqref{eq52} can be rewritten as
\begin{equation}
\begin{split}
{\rm MSE}\{\hat{\bf f}_k\}_2 & =E[{\bf\xi}^H(P_{s,k}{\bf R}_{{\bf
f}_k}+P_{s,k}{\bf R}_{{\bf e}_k}+P_n){\bf P}_k (P_{s,k}{\bf R}_{{\bf
f}_k} \\ &\quad +P_{s,k}{\bf R}_{{\bf e}_k}+P_n){\bf\xi}] -2E[{\bf
f}_k^H(P_{s,k}{\bf R}_{{\bf f}_k} +P_{s,k}{\bf R}_{{\bf e}_k}\\ &
\quad +P_n){\bf\xi}] +E[{\bf f}_k^H{\bf f}_k]. \label{eq53}
\end{split}
\end{equation}
After expanding the terms with the multiplications, eliminating the
uncorrelated ones and taking into account that ${\bf f}_k$ is
normalized, i.e., $E[{\bf f}_k^H{\bf f}_k]=1$, we obtain
\begin{equation}
\begin{split}
{\rm MSE}\{\hat{\bf f}_k\}_2 & = E[{\bf\xi}^H(P^2_{s,k}{\bf R}_{{\bf e}_k}{\bf P}_k{\bf R}_{{\bf e}_k}+P_nP_{s,k}{\bf P}_k{\bf R}_{{\bf e}_k}   \\
& \quad +P_n^2{\bf P}_k){\bf\xi}+1].  \label{eq54}
\end{split}
\end{equation}
By substituting \eqref{eq21} in \eqref{eq54} and performing further
simplifications, we get
\begin{equation}
\begin{split}
{\rm MSE}\{\hat{\bf f}_k\}_2 & =E[(P^2_{s,k}\epsilon^2||{\bf R}_{{\bf f}_k}||^2_F+P_nP_{s,k}\epsilon||{\bf R}_{{\bf f}_k}||_F+P_n^2) \\
& \quad {\bf\xi}^H{\bf P}_k{\bf\xi}+{\bf f}_k^H{\bf f}_k] \\
& =(\frac{1}{3}P^2_{s,k}\epsilon^2_{max}E[||{\bf R}_{{\bf
f}_k}||_F]^2+\frac{1}{2}P_nP_{s,k}\epsilon_{max}E[||{\bf R}_{{\bf
f}_k}||_F]\\ & \quad  +P_n^2) {\bf\xi}^HE[{\bf P}_k]{\bf\xi}+1,
\label{eq55}
\end{split}
\end{equation}
which monotonically increases with respect to $E[||{\bf R}_{{\bf
f}_k}||_F]$. However, the results of the analysis can become more
insightful if we compare the MSE obtained in the two approaches
considered in this section. Let us denote them as ${\rm
MSE}\{\hat{\bf f}_k\}_1$ (described in \eqref{eq35}) and ${\rm
MSE}\{\hat{\bf f}_k\}_2$ (described in \eqref{eq55}), respectively,
and ${\bf\xi}^HE[{\bf P}_k]{\bf\xi}$ as $\tau$. If we compute their
difference we have
\begin{equation}
\begin{split}
{\rm MSE}\{\hat{\bf f}_k\}_2-{\rm MSE}\{\hat{\bf f}_k\}_1 & =
\bigg(\frac{1}{3}P^2_{s,k}\epsilon^2_{max}E[||{\bf R}_{{\bf
f}_k}||_F]^2
\\  & \quad +\frac{1}{2}P_nP_{s,k}\epsilon_{max}E[||{\bf R}_{{\bf
f}_k}||_F]+P_n^2\bigg)\tau \\ & \quad
+1-\frac{M}{2}\epsilon_{max}E[||{\bf R}_{{\bf f}_k}||_F].
\label{eq56}
\end{split}
\end{equation}
If we take the partial derivative of \eqref{eq56} with respect to
$\tau$, then we have
\begin{equation}
\frac{\partial\{{\rm MSE}\{\hat{\bf
f}_k\}_2-{\rm MSE}\{\hat{\bf f}_k\}_1\}}{\partial{\tau}}> 0, \label{eq57}
\end{equation}
which implies that ${\rm MSE}\{\hat{\bf f}_k\}$ is proportionally
and monotonically increasing with respect to $\tau$. From
\eqref{eq40} we have $${\rm max}\{||{\bf R}_{{\bf f}_k}||_F\} <
\sqrt{M\lambda_{max,k}^2-2\sigma_{\lambda,k}\lambda_{max,k}+\sigma_{\lambda,k}^2}<\sqrt{M}\lambda_{max,k},$$
which yields
\begin{equation}
\begin{split}
{\rm MSE}\{\hat{\bf f}_k\}_2-{\rm MSE}\{\hat{\bf f}_k\}_1 &
<\bigg(\frac{1}{3}P^2_{s,k}\epsilon^2_{max}M\lambda_{max,k}^2 \\
& \quad +\frac{1}{2}P_nP_{s,k}\epsilon_{max}\sqrt{M}\lambda_{max,k}+P_n^2 \bigg)\tau \\
& \quad +1-\frac{M}{2}\epsilon_{max}\sqrt{M}\lambda_{max,k}.
\label{eq58}
\end{split}
\end{equation}
In other words, if the right-hand side of \eqref{eq58} is less than
$0$ when $\tau$ satisfies
\begin{equation}
\tau <
\frac{\frac{M}{2}\epsilon_{max}\sqrt{M}\lambda_{max,k}-1}{\frac{1}{3}P^2_{s,k}\epsilon^2_{max}M\lambda_{max,k}^2
+\frac{1}{2}P_nP_{s,k}\epsilon_{max}\sqrt{M}\lambda_{max,k}+P_n^2},
\label{eq59}
\end{equation}
and $${\rm MSE}\{\hat{\bf f}_k\}_2-{\rm MSE}\{\hat{\bf f}_k\}_1 <
0$$ is true for all possible values of $E[||{\bf R}_{{\bf
f}_k}||_F]$, which indicates a smaller MSE result from approach 2
(${\rm MSE}\{\hat{\bf f}_k\}_2$) as compared to approach 1 (${\rm
MSE}\{\hat{\bf f}_k\}_1$). Interestingly, this indicates that using
prior knowledge about the mismatch in the form of low-rank subspace
and cross-correlation processing can result in smaller values of
MSE. However, the only term of $\tau$ that has to be determined is
the subspace projection matrix ${\bf P}_k$, which is dependent on
its subspace properties and can be further exploited by eigenvalue interpolation methods \cite{r18}.

Based on \cite{r19} and assuming that the channels and input data have Gaussian distribution, the MMSE and SINR of the system are related by
\begin{equation}
    {\rm MMSE}\{z\}=\frac{1}{1+{\rm SINR}_{max}}, \label{eq60}
\end{equation}
where the ${\rm SINR}_{max}$ is obtained in \eqref{eq34}. For simplicity, let us drop the time index $i$ and denote the term $(P_n{\bf I}_M +\hat{\bf D}^{-1/2}\hat{\bf U}\hat{\bf D}^{-1/2})^{-1} \hat{\bf D}^{-1/2}\hat{\bf R}_1\hat{\bf D}^{-1/2}$ in \eqref{eq34} and \eqref{eq33} as
\begin{equation}
  {\bf\Theta} = (P_n{\bf I}_M +\hat{\bf D}^{-1/2}\hat{\bf U}\hat{\bf D}^{-1/2})^{-1} \hat{\bf D}^{-1/2}\hat{\bf R}_1\hat{\bf D}^{-1/2}. \label{eq_def}
\end{equation}
According to the properties of eigenvalues and eigenvectors, we know that
\begin{equation}
{\bf\Theta}{\mathcal P}\{ \bf\Theta \}=\lambda_{largest}\{ \bf\Theta \}{\mathcal P}\{ \bf\Theta \}. \label{eq61}
\end{equation}
If we pre-multiply both sides of \eqref{eq33} by $\hat{\bf D}^{1/2}$ and drop the time index, it becomes
\begin{equation}
    \frac{\hat{\bf D}^{1/2}{\bf w}}{\sqrt P_T} = {\mathcal P}\{ \bf\Theta \}. \label{eq62}
\end{equation}
Then, by substituting \eqref{eq62} in \eqref{eq61}, we obtain
\begin{equation}
    {\bf\Theta}\hat{\bf D}^{1/2}{\bf w} = \lambda_{largest}\{\bf\Theta\}\hat{\bf D}^{1/2}{\bf w}, \label{eq63}
\end{equation}
which through the use of the weight vector norm constraint ${\bf w}^H{\bf w}=1$ gives the expression for $\lambda_{largest}\{\bf\Theta\}$:
\begin{equation}
    \lambda_{largest}\{\bf\Theta\}={\bf w}^H\hat{\bf D}^{-1/2}{\bf\Theta}\hat{\bf D}^{1/2}{\bf w}. \label{eq64}
\end{equation}
Finally, by substituting \eqref{eq64} in \eqref{eq34}, ${\rm SINR}_{max}$ is rewritten as
\begin{equation}
    {\rm SINR}_{max}=P_T\lambda_{largest}\{\bf\Theta\}=P_T{\bf w}^H\hat{\bf D}^{-1/2}{\bf\Theta}\hat{\bf D}^{1/2}{\bf w}
    \label{eq65}
\end{equation}
while the MMSE of $z$ is computed from \eqref{eq60} as
\begin{equation}
    {\rm MMSE}\{z\}=\frac{1}{1+P_T{\bf w}^H\hat{\bf D}^{-1/2}{\bf\Theta}\hat{\bf D}^{1/2}{\bf w}},
    \label{eq66}
\end{equation}
In \eqref{eq66}, ${\rm MMSE}\{z\}$ is determined by ${\bf w}$ and ${\bf\Theta}$, which are both directly obtained from \eqref{eq33} and \eqref{eq_def} and are only dependent on the variables of the channel estimates $\hat{\bf f}_k$, $k=1,\dotsb,K$ and ${\bf g}$. We remark that the weight vector ${\bf w}$ (or its diagonal matrix form ${\bf W}$) is estimated and expressed using $\hat{\bf f}_k$, $k=1,\dotsb,K$ and ${\bf g}$, as we know that $\hat{\bf D}$, $\hat{\bf R}_1$, $\hat{\bf U}$ are all based on $\hat{\bf f}_k$, $k=1,\dotsb,K$ and ${\bf g}$, where the MMSE of ${\bf g}$ can be obtained in a similar way as $\hat{\bf f}_k$ and hence the derivation is omitted. Therefore, the MMSE of $z$ for the proposed LRCC-RDB technique only depends on the MMSE estimates for $\hat{\bf f}_k$ and $\hat{\bf g}$. In this case, ${\rm MMSE}\{ z \}$ is proportional to ${\rm MMSE}\{ \hat{\bf f}_k \}$ and ${\rm MMSE}\{ \hat{\bf g} \}$, i.e., ${\rm MMSE}\{ z \} \propto {\rm MMSE}\{ \hat{\bf f}_k \}, {\rm MMSE}\{ \hat{\bf g} \}$. Since we have proved that under certain assumptions the estimate $\hat{\bf f}_k$ obtained by the proposed LRCC-RDB approach is better than those of other analyzed techniques then the same analysis procedure applies to $\hat{\bf g}$. Therefore, RDB techniques which adopt the LRCC-RDB technique are able to achieve smaller ${\rm MMSE}\{ \hat{\bf f}_k \}$ and ${\rm MMSE}\{ \hat{\bf g} \}$, and consequently a smaller ${\rm MMSE}\{ z \}$ than those of other RDB methods, as verified in the simulation results.

\subsection{Complexity Analysis}

This subsection presents an analysis of the computational complexity
of LRCC-RDB and comparisons with existing robust
approaches such as the most common worst-case approaches, which are
typically solved using an interior point method (e.g.
\cite{r11,r14,r21,r20}) and the probabilistic based stochastic
approach introduced in \cite{r9}, all of which are shown in Table
\ref{table2}. It should be noted that we compare the computational
complexities of the approaches regardless of their system design
objectives, which could target the minimization of the total relay
transmit power with a QoS (output SNR or SINR) constraint (e.g.
\cite{r11,r14,r21}). A limited number of techniques like the
approach reported in \cite{r20} aims to maximize the QoS with
individual relay transmit power, whereas LRCC-RDB
aims to maximize the output SINR with a total relay transmit power constraint and does not require any online optimization
procedure. Due to the highly involved procedures and recursions of
online convex optimization employed by the existing algorithms, we
only compare the polynomial bound of each of them. Note that LRCC-RDB only incurs cubic complexity (${\mathcal O}(M^3)$) when computing the weight vector. Specifically, $\hat{\bf
D}^{-1/2}(i)$ is a diagonal matrix, which means the computations of $\hat{\bf
D}^{-1/2}(i)\hat{\bf U}(i)\hat{\bf D}^{-1/2}(i)$ and $\hat{\bf
D}^{-1/2}(i)\hat{\bf R}_1(i)\hat{\bf D}^{-1/2}(i)$ are both at a cost of ${\mathcal O}(M^2)$. The only costly operations are: the matrix inversion $(P_n{\bf I}_M+\hat{\bf D}^{-1/2}(i)\hat{\bf U}(i)\hat{\bf D}^{-1/2}(i))^{-1}$, the matrix multiplication between $(P_n{\bf I}_M+\hat{\bf D}^{-1/2}(i)\hat{\bf U}(i)\hat{\bf D}^{-1/2}(i))^{-1})$ and $(\hat{\bf D}^{-1/2}(i)\hat{\bf R}_1(i)\hat{\bf D}^{-1/2}(i))$, the eigen decomposition of the resulting matrix after multiplication. Each of these three operations requires a complexity of ${\mathcal O}(M^{3})$ and only needs to be computed once. As previously explained, we only need to compute the above required operations once for a data block when the channel is within the same coherence time, that is, only one matrix inversion, one matrix multiplication and one eigen decomposition are required for a data block. For SOCP and SDP, there are multiple iterations within the solver per snapshot as well. The key point is that either per snapshot or when computing the weights once in a data block, the proposed approach is computationally simpler and provides improved resilience and robustness.

\begin{table}[htb!]
\small
\begin{center}
\caption{Computational Complexity}
\begin{tabular}{|c|c|}
\hline
RDB algorithms & Flops per snapshot \\
\hline
Worst-case with SDP \cite{r11,r14} & ${\mathcal O}(M^{6.5})$ \\
\hline
Worst-case with SOCP \cite{r21,r20} & ${\mathcal O}(M^{3.5})$ \\
\hline
Stochastic approach with SDP \cite{r9} & ${\mathcal O}(M^{9.5})$ \\
\hline
Proposed LRCC-RDB method & ${\mathcal O}(M^3)$ \\
\hline
\end{tabular} \label{table2}
\end{center}
\end{table}

\section{Simulations}

\begin{figure}[!htb]
\begin{center}
\def\epsfsize#1#2{0.95\columnwidth}
\epsfbox{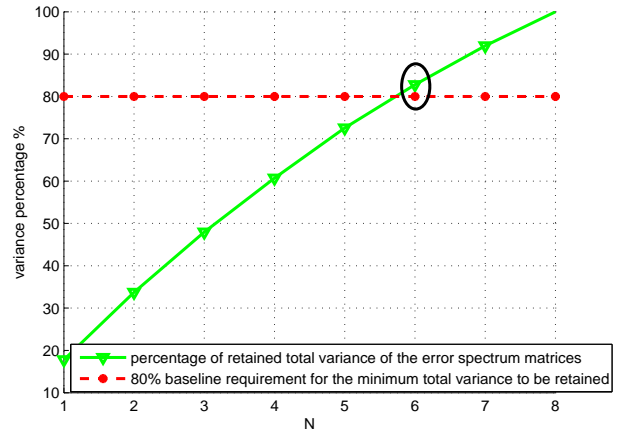} \caption{Principal components selection}
\label{pc_selection}
\end{center}
\end{figure}

In this section we conduct simulations to assess the proposed
LRCC-RDB method for several scenarios, namely, the case of perfect
CSI, the case in which no robust method is used and the CSI is
imperfect, and the cases in which CSI is imperfect and several
existing robust approaches
\cite{r13,r12,r11,r25,r14,r21,r26,r15,r24,r20} (i.e. worst-case SDP
online programming) are used. The figures of metric considered
include the system output SINR versus input SNR as well as the
maximum allowable total transmit power $P_T$. We also examine
incoherent scenarios, where some of the interferers are strong
enough as compared to the desired signal and the noise. In all
simulations, the system input SNR is known and can be controlled by
adjusting only the noise power. Both channels $\hat{\bf F}$
and $\hat{\bf g}$ are modeled by the Rayleigh distribution. The
shadowing and path loss effects are taken into account with the path
loss exponent set to $\rho=2$, the source-to-destination power path
loss set to $L=10$dB and the shadowing spread set to $\sigma_s=3$dB.
As discussed in Section II, the relative source-to-relay link distances $d_{s,r_m}$ are selected from a set of uniform random variables distributed between $0.5$ to $0.9$, with
corresponding relay-source-destination angles $\theta_{r_m,s,d}$
randomly chosen from an angular range between $-\pi/2$ and $\pi/2$.
The total number of relays and signal sources are set to $M=8$ and
$K=3$, respectively, and we set $\sigma_{b_k}=1$ for $k=1,\dotsb,K$.
The interference-to-noise ratio (INR) of the system is fixed at
$10$dB unless otherwise specified. A total number of $100$ snapshots are considered. The number of principal components is selected according to the criterion mentioned in Section III to optimize the performance of LRCC-RDB,  as illustrated in Fig. \ref{pc_selection}. Assigning any value larger than the minimum sufficient value of $N$ leads to insignificant performance improvements and extra complexity.

\begin{figure}[!htb]
\begin{center}
\def\epsfsize#1#2{0.95\columnwidth}
\epsfbox{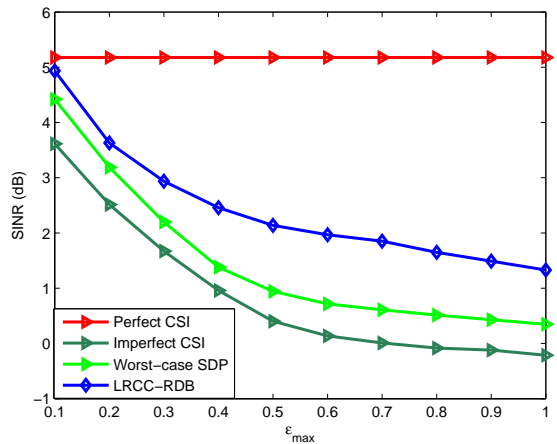} \caption{SINR versus $\epsilon_{max}$,
$P_T=1$dBW, SNR=$10$dB, INR=$10$dB} \label{figure-eps}
\end{center}
\end{figure}

In the first example, we examine the SINR performance against
different values of the mismatch parameter $\epsilon_{max}$
($0.1\leq\epsilon_{max}\leq1$) in Fig. \ref{figure-eps}, while
limiting the maximum allowable transmit power to $P_T=1$dBW and
fixing the input SNR at $10$dB for all the compared cases. The
powers of the interferers are equally distributed across the
interferers. The worst-case SDP method is adopted from \cite{r11},
in which the values of $\epsilon_{max}$ are set to be consistent with
all the mismatched matrix quantities. The results show that the
proposed LRCC-RDB method preserves the robustness against the
increase of the level of channel errors and remains close to the
case of perfect CSI, whereas the worst-case SDP method suffers
performance degradation against the increase of the level of channel
errors.

\begin{figure}[!htb]
\begin{center}
\def\epsfsize#1#2{0.95\columnwidth}
\epsfbox{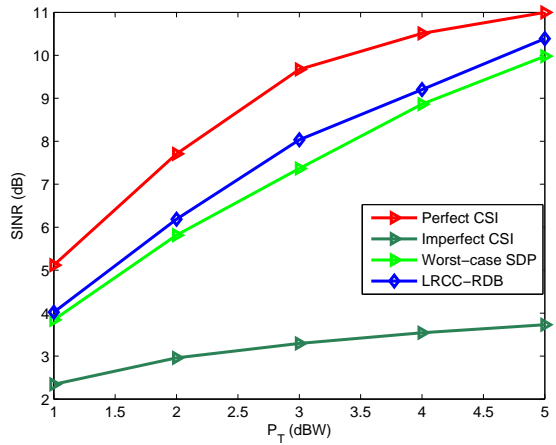}
\caption{SINR versus $P_T$, SNR=$10$dB, $\epsilon_{max}=0.5$, INR=$10$dB} \label{figure4}
\end{center}
\end{figure}

In the second example, we examine the SINR performance versus the
variation of maximum allowable total transmit power $P_T$ (i.e.
$1$dBW to $5$dBW) by fixing the input SNR at $10$dB. We consider the
same INR and that all interferers have the same power. In this
example, we set the perturbation parameter to $\epsilon_{max}=0.5$
for all compared techniques. In Fig. \ref{figure4}, it shows the
output SINR increases as we increase the limit for the maximum
allowable transmit power and this results in a substantial
difference when a robust approach is used. LRCC-RDB
outperforms the worst-case SDP algorithm and performs close
to the case with perfect CSI.

\begin{figure}[!htb]
\begin{center}
\def\epsfsize#1#2{0.95\columnwidth}
\epsfbox{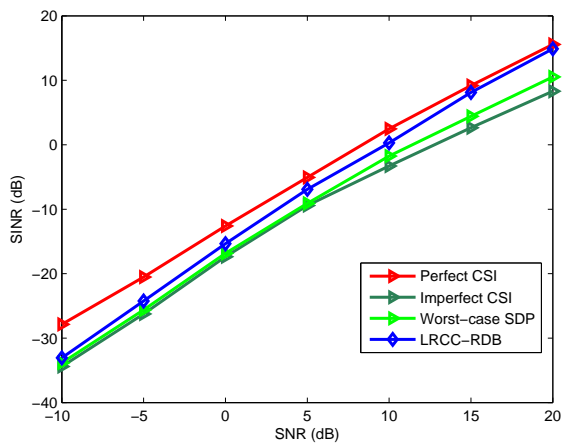}
\caption{SINR versus SNR, $P_T=1$dBW, $\epsilon_{max}=0.2$, INR=$20$dB} \label{figure5}
\end{center}
\end{figure}

\begin{figure}[!htb]
\begin{center}
\def\epsfsize#1#2{0.95\columnwidth}
\epsfbox{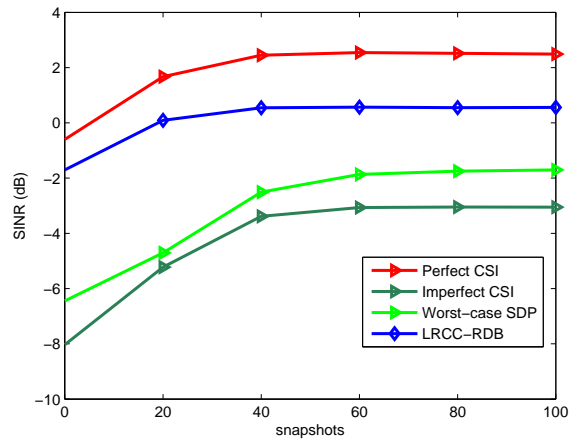} \caption{SINR versus snapshots, $P_T=1$dBW,
$\epsilon_{max}=0.2$, SNR=$10$dB, INR=$20$dB} \label{figure6}
\end{center}
\end{figure}

In the last example, we increase the system INR from $10$dB to
$20$dB. We consider $K=3$ users (which means there are two
interferers in total) but rearrange the powers of the interferers so
that one of them is much stronger than the other. Specifically, we
examine the compared approaches in an incoherent scenario and set
the power ratio of the stronger interferer over the weaker one to
$10$. The maximum allowable total transmit power $P_T$ and the
perturbation parameter $\epsilon_{max}$ are set to $1$dBW and $0.2$,
respectively. We observe the SINR performance versus SNR for these
techniques and illustrate the results in Fig. \ref{figure5}. Then we
set the system SNR to $10$dB and observe the output SINR performance
versus snapshots as in Fig. \ref{figure6}. It can be seen that all
the approaches have performance degradations due to the strong
interferers as well as their power distribution. However, LRCC-RDB shows robustness in terms of output SINR
performance against the presence of strong interferers with
unbalanced power distribution. In particular, with relative high
system SNRs, LRCC-RDB is able to perform extremely close
to the case of perfect CSI.

\section{Conclusion}

We have devised a novel RDB approach based on the exploitation of
the cross-correlation between the received data from the relays and
the system output, as well as a low-rank subspace projection method
to estimate the channel errors. In the proposed LRCC-RDB method, a
total relay transmit power constraint has been considered and the
objective is to maximize the output SINR. A performance analysis of
LRCC-RDB has been carried out and shown that it outperforms
approaches that do not exploit prior knowledge about the mismatch.
LRCC-RDB does not require any costly online optimization procedure
and the simulation results have shown excellent performance as
compared to existing approaches and can be applied to detection and
estimation in wireless communications
\cite{spa,mfsic,gmibd,tdr,mbdf,mmimo,lsmimo,armo,siprec,did,lclrbd,gbd,wlbd,bbprec,mbthp,rmbthp,badstbc,bfidd,1bitidd,baplnc,jpbnet}.

\bibliographystyle{IEEEtran}
\bibliography{bibliography}

\end{document}